\documentclass[conference]{IEEEtran}
\IEEEoverridecommandlockouts
\usepackage{cite}
\usepackage{amsmath,amssymb,amsfonts}
\usepackage{algorithmic}
\usepackage{graphicx}
\usepackage{textcomp}
\usepackage{xcolor}
\usepackage{subcaption}
\usepackage{relsize}

\usepackage{amsmath,amssymb}

\usepackage{booktabs, multirow} 
\usepackage{soul}
\usepackage{colortbl}
\usepackage{changepage,threeparttable} 

\usepackage[linesnumbered,ruled,vlined]{algorithm2e}
\SetCommentSty{mycommfont}
\SetKwInput{KwInput}{Input}                
\SetKwInput{KwOutput}{Output}              
\def\BibTeX{{\rm B\kern-.05em{\sc i\kern-.025em b}\kern-.08em
    T\kern-.1667em\lower.7ex\hbox{E}\kern-.125emX}}
\begin{document}

\title{Towards Detecting IoT Event Spoofing Attacks Using Time-Series Classification\\
\thanks{The work has been supported by the Cyber Security Research Centre Limited, whose activities are partially funded by the Australian Government’s Cooperative Research Centres Programme.}
}


\author{
    \IEEEauthorblockN{Uzma Maroof\IEEEauthorrefmark{1}\IEEEauthorrefmark{2}\IEEEauthorrefmark{3}, Gustavo Batista \IEEEauthorrefmark{3}, Arash Shaghaghi\IEEEauthorrefmark{3}, and Sanjay Jha\IEEEauthorrefmark{2}\IEEEauthorrefmark{3}}
    \IEEEauthorblockA{\IEEEauthorrefmark{1}University of Waterloo, Waterloo, Canada}
    \IEEEauthorblockA{\IEEEauthorrefmark{2}Cybersecurity CRC, Australia} 
    \IEEEauthorblockA{\IEEEauthorrefmark{3}The University of New South Wales, Sydney, Australia}
    }

\maketitle

\maketitle

\begin{abstract}
Internet of Things (IoT) devices have grown in popularity since they can directly interact with the real world. Home automation systems automate these interactions. IoT events are crucial to these systems' decision-making but are often unreliable. Security vulnerabilities allow attackers to impersonate events. Using statistical machine learning, IoT event fingerprints from deployed sensors have been used to detect spoofed events. Multivariate temporal data from these sensors has structural and temporal properties that statistical machine learning cannot learn. These schemes' accuracy depends on the knowledge base; the larger, the more accurate. However, the lack of huge datasets with enough samples of each IoT event in the nascent field of IoT can be a bottleneck. In this work, we deployed advanced machine learning to detect event-spoofing assaults. The temporal nature of sensor data lets us discover important patterns with fewer events. Our rigorous investigation of a publicly available real-world dataset indicates that our time-series-based solution technique learns temporal features from sensor data faster than earlier work, even with a 100- or 500-fold smaller training sample, making it a realistic IoT solution.

\end{abstract}


\section{Introduction}\label{c3_Intro}

IoT devices are expected to be the cornerstone for the next generation of ubiquitous computing \cite{gubbi2013internet}, constantly transforming daily life. Global IoT device installations will reach 30.9 billion by 2025. \cite{vailsheryiot}. Several factors contribute to this increased popularity, but the variety of possible interactions between various devices may be the most essential. These interactions may involve other devices, remote users, or cloud services like Amazon Web Services, Google Cloud Platform, Instagram, etc. A striking aspect of these devices that distinguishes them from other computer devices (desktops/laptops/cell phones) is their capability to interact with the physical world. For instance, a smart humidifier may activate when humidity drops below a threshold. Likewise, a smart lock might unlock once the homeowner arrives at the front door.



End-users previously configured these smart home devices through smartphone applications. As the number of deployed devices grows, manual configuration becomes increasingly cumbersome. Moreover, these mobile apps typically do not support cross-vendor device communications.

\par To solve this problem, home-automation platforms have emerged to allow for the easy connection of devices from different manufacturers and the wide personalization of IoT rollouts. These platforms rapidly became popular due to their standardized device abstractions and simple configuration interfaces. Samsung's SmartThings, Apple's HomeKit, openHAB, and Microsoft Flow are popular IoT platforms. 
Typically, these systems use the \textbf{Trigger-Action Paradigm (TAP)} to enable end-users to establish device interactions using simple rules, resulting in novel functionalities. Upon receiving a \textbf{\textit{trigger event}} (e.g., user approaching home), the device must execute the \textbf{\textit{action(s) event}} (e.g., heating system activation) as per the TAP rule. A recent study indicated that TAP programming could support 80\% of end-user customization due to its conceptual simplicity. This point is further strengthened by reports indicating that millions of user installations execute billions of rules monthly \cite{ martin2019ifttt}. 


While these platforms offer convenience, they also create vulnerabilities that are easier to exploit and enable new attacks. For example, a momentary burst of light has the ability to initiate malware on a device \cite{sikder20176thsense}. These inexpensive devices can be used by attackers to execute \textbf{\textit{Event-spoofing attacks}}, generating fake trigger events and triggering needless automation rules. A hacked IoT device can fake a successful completion of an action event, even if it blocks its execution.


Exploiting these security vulnerabilities has dual repercussions. First, IoT devices, tightly intertwined with their surroundings, can spread attacks to the real world. Second, because these home automation rules are linked, an attacker with control of one device can utilize them as part of a complicated attack chain to take over numerous devices by abusing TAP rules.

Much research has focused on IoT vulnerabilities; however, home automation vulnerabilities have recently gained attention. Most IoT security solutions are tailored from generic computing device paradigms rather than IoT-specific requirements, which is why they fail. For example, to mitigate compromised devices, some existing literature has focused on traditional methods of analyzing source/binary code, event logs, and installation rules \cite{hsu2019safechain, gu2020iotgaze, fu2021hawatcher, zhang2018homonit}. However, the complexity of these schemes grows exponentially with the number of connected devices and rules in place. Other studies focus on real-time IoT event verification to find compromised devices. Triggered events and actions are verified before firing rules and consequently updating the system state. Commonly, IoT event fingerprints are built using wireless network features \cite{gu2020iotgaze,acar2020peek}.


Birnbach et al.\cite{birnbach2019peeves} took a different method and exploited IoT devices' impact on the physical environment. Their system, called PEEVES, learns event fingerprints from nearby sensors to verify IoT events. IoT event fingerprints were learned using statistical feature extraction machine learning.

\begin{figure}

\centering
    \includegraphics[width=0.8\linewidth]{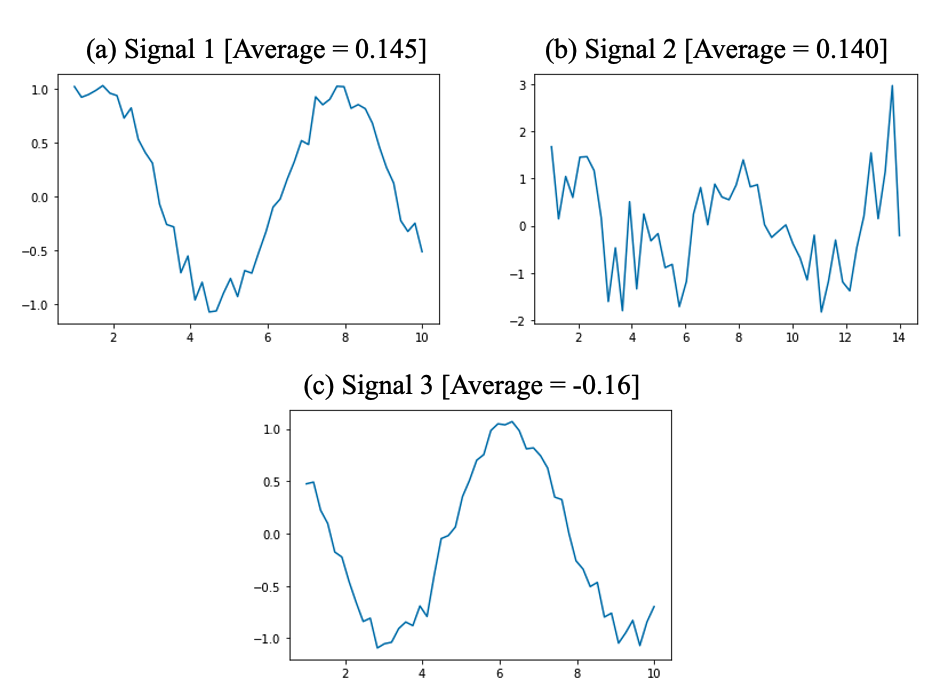}
    \caption{Statistical feature such as average signal strength unable to recognise that Signal 3 is a shifted version of Signal 1.}\label{Fig_tsc_signals}
\end{figure}

PEEVES, a multivariate time-series dataset, records diverse sensor signals with key temporal/ordinal characteristics. Statistical feature extraction fails to capture these temporal aspects, resulting in information loss. Consider a simple example as shown in figure \ref{Fig_tsc_signals} comparing three time-series signals in. Signal 3 is a slightly distorted and displaced version of Signal 1, while Signal 2 is unrelated to the other two. The average signal strength of the three signals yields 0.145, 0.140, and -0.16 for Signals 1, 2, and 3, respectively. Despite their similarities, Signals 1 and 3 have drastically different average values. Interestingly, Signals 1 and 2 average similarly. In this simple example, statistical characteristics fail to capture time-series signal structure and temporal patterns. Thus, statistical classifiers perform poorly. 




To compensate for their inability to maintain temporal properties that result in information loss, statistical machine learning systems rely on large datasets. The IoT ecosystem is enormously heterogeneous, making such approaches impractical. New devices and events are introduced often. Multiple factors cause this dynamicity, such as: (1) permanent addition or removal of devices, such as installing an additional smart lock. (2) Temporarily addition or removal of devices, such as passengers entering or exiting a smart-airport, may add/remove wearables numerous times in a day. (3) Adding features, such as Bluetooth connectivity, to an existing gadget. (4) The consumer upgrades to a different category of equipment, such as a smart vacuum cleaner. In this nascent field of IoT, the lack of massive datasets can be a huge bottleneck. Hence, a large and stable knowledge base with enough occurrences of each IoT event is unlikely.

Thus, event-spoofing detection systems must account for the heterogeneous and dynamic IoT world and not require enormous datasets to achieve minimum accuracy. This paper introduces an event-spoofing system that extracts structural features and retains signal temporal properties. We demonstrate that these methods use a significantly smaller dataset than statistical-feature-based methods. Our method uses Dynamic Time Warping (DTW) similarity metric (refer to section \ref{c3_TSC_DisMsr}), which is capable of signal structural comparison \cite{duin2012dissimilarity}. For instance, the DTW distance between Signals 1 and 3 is 1.46, indicating closeness. The DTW distance between Signals 2 \& 1 (6.89) and Signals 2 \& 3 (7.46) indicates accurate proximity measurement. Clearly, different signals have large distances. Numerous studies indicate that dissimilarity space techniques can outperform classifiers that directly operate on original space\cite{bunke2008graph,riesen2007graph}. This method combines the advantages of structural representation and statistical learning.

This paper detects IoT event-spoofing attacks using TSC. Our thesis is that a minimum knowledge base should be enough to create a stable learning model using TSC-based classification algorithms that preserve dataset temporal features, thus, it should be suitable for IoT paradigm. The main contributions of this research are:
\begin{enumerate}
    \item We propose using time-series classification to extract structural characteristics from IoT event data, enabling successful verification with minimum training data (section \ref{c3_TSC}).
    \item We design and implement a complete IoT event detection system based on Dynamic Time Warping to extract structural features (section \ref{c3_Proposed_System}).
    \item Our systematic evaluation demonstrates the effectiveness and efficiency of our proposed system, achieving equivalent detection accuracy with a 100-500 times smaller training dataset (see section \ref{c3_Evaluation}).
    \item Our results reflect real-world use scenarios by using PEEVES \cite{birnbach2019peeves} instead of synthetical training and evaluation data.
    \item We also made our solution public for further research investigation.


\end{enumerate}


\section{Background \label{c3_background}}
In this study, we used Birnbach et al.'s time-series dataset \cite{birnbach2019peeves}. As discussed below, due to the various data particularities, it is a complex classification problem. This section briefly describes the dataset and data particularities that lead to various challenges in designing a TSC solution. 


The PEEVES dataset tracks four smart environment users for 13 days. It depicts a smart workplace by tracking 22 IoT events from ten devices. For instance, a smart door can generate "Door open" and "Door close" events, whereas a smart fan can generate "Fan on" and "Fan off" events. Each IoT event is a tuple: $\langle Timestamp, Event \rangle$.

   


where $Event$ is a binary event occurrence value. The location of a reported event is stored as 1, while a 0 is saved every second afterwards. Numerous sensors sensing 12 sensor-modality categories measure IoT events' impact on physical media. There are 42 sensors on 12 Raspberry Pi devices. The Raspberry Pi device stores numerous sensor readings each second. Network Time Protocol syncs Raspberry Pi device times. Each Sensor stores data as tuples: $\langle Timestamp, Sensor Reading  \rangle$


Some sensors, such as light sensors, record thousands of values per second, thus very high-resolution Timestamp data is used.
\section{Challenges}\label{c3_challenges}
We now discuss some unique aspects of the dataset that made it challenging to apply any learning algorithms. 
\begin{itemize}
    \item \textbf{\textit{Multi-variate time-series}}:
    
    The dataset exhibits the impact of 22 IoT events on physical media, recorded by an array of 235 time-series signals. Thus, it can be categorized as a complex, high-dimensional, multi-variate time-series problem.

    \item \textbf{\textit{Variable series length}}: 
    The frequencies of these time-series signals vary greatly. While a power meter sensor records one or two values per second, a temperature sensor may record twenty. Moreover, sensors of the same type may record different values for the same period, resulting in unequal signal durations. Apart from high computational cost, truncating or padding all these signals will result in information loss or noise addition.
    \item \textbf{\textit{Unbalanced dataset}} 
     The presence of IoT events is marked as a 1-event and absence as a 0-event. However, IoT events may be rare.  For instance, users who avoid caffeine may not record "Coffee-Machine Used" events. Thus, only 2773 1-events were gathered for 22 IoT events throughout 13 days. However, by sampling "no-event occurred" every second, over a million 0-events can be captured. This enormously unbalanced dataset will bias any time-series classifier towards 0-events, resulting in low classification accuracy. 


    \item \textbf{\textit{Computational Complexity}}: 
    In the dataset, 235 sensors recorded four users' daily activities during 13 days, generating 300 GB of data with over a million incidents. Many TSCs transform data to a different space, but doing so on such an enormous dataset can cause performance issues.
    
    

\end{itemize}

\section{Time-series Classification(TSC) \& Design decisions \label{c3_TSC}}
Classification of time-series data is a popular topic due to its rapid applications in many disciplines. Various TSC approaches have been developed for different problem domains. Due to our problem's intricacy, TSC methods had to be carefully selected. Following a background on TSC, we discuss TSC methods.

\textbf{Definition 1}\textbf{ \textit{Univariate time-series}} is a sequence of measurements taken at regular intervals in time. Univariate time-series has only one independent variable and is denoted as:
\begin{equation*}
    T= (t_1,t_2, \dots t_n)
\end{equation*}

\textbf{Definition 2} \textbf{\textit{Multivariate time-series} }is a collection of time-series that share the same timestamps. Two univariate time-series $T1=(t1_1,t1_2, \dots t1_n)$ and $T2=(t2_1,t2_2, \dots t2_n)$ can be combined into a bivariate time-series $T_B$ as $((t1_1,t2_1),(t1_2,t2_2), \dots (t1_n,t2_n))$. This approach can be extended to a multivariate series $T_M$ comprising $m$ variables and $n$ intervals. It can be represented as an $m\times n$ matrix $M$, where $M_{i,j}$ represents the value of the $j^{th}$ univariate time-series at time $i$.

\textbf{Definition 3} \textbf{\textit{Time-series classification}} is a supervised machine learning method that predicts a \textbf{\textit{label}} from a time-series input. Here, discovering discriminative features from the time-series to distinguish the classes is the major challenge. Formally, TSC is a mapping from a time-series $T$ to a finite collection of class labels $C = \{c_1 \dots c_{|C|}\}$, which associates each element of $T$ to a label from $C$. Thus, $f(\cdot ): T \longrightarrow C$

    


We look at why raw time-series data is unsuitable for typical machine learning analysis \cite{renard2017time}. Consider the input time-series $T$ in $\mathbb{R}^m$ as a feature vector of $m$ real-valued random variables. The correlation between successive points in time, i.e., the order of $T$'s random variables, is critical. For $T$ dimensions, all dataset instances should have the same distribution. Thus, all dataset time-series must align. However, Possible time-series distortions make this condition difficult to meet:


\begin{enumerate}
    \item Time-series instances may vary in length across the dataset. Thus, the resulting feature vectors will vary in dimension. Changing input size on the fly in machine-learning systems is difficult.
    \item Time-series tracks structural patterns, and it is challenging to pinpoint their start and end. A pattern can occur at many frequencies with occasional accelerations, decelerations, and gaps. 
    \item Multivariate time-series reinforce the above concerns.

\end{enumerate}

Many methods have been developed to overcome the limitations of applying traditional machine learning classifiers to time-series data. These methods address two key difficulties. \cite{susto2018time}:
\begin{enumerate}
    \item How to compare time-series with varying durations?
    \item Which time-series are representations of a class phenomenon?
\end{enumerate}
We now discuss various TSC methods and design choices related to our problem.

\subsection{Time Series Classification methods}
TSC methods are broadly classified as feature-based, model-based, and distance-based methods. The \textbf{\textit{Feature-based methods}} aims to turn temporal problems into conventional ones by extracting static information, such as mean and standard deviation, global frequency content, or motif discovery. Alternatively, \textbf{\textit{Model-based methods}} such as the Hidden Markov Model (HMM), create a system model and classify each series based on its best fit. Lastly, \textbf{\textit{distance-based methods}} quantify dissimilarities between series and integrate them into traditional learning methods. Since, autonomous feature extraction and model construction can be challenging with the first two methods, therefore, we investigated distance-based TSC due to its lower complexity \cite{abanda2019review}.



\subsection{Dissimilarity measures \label{c3_TSC_DisMsr}}
A suitable dissimilarity metric is required for Distance-based TSC. 
Lock-step and elastic dissimilarity measures are two primary categories.
\textbf{Lock-step measures}, such as \textbf{\textit{Euclidean distance(ED)}}, compare time-series point-by-point. $ED$ between two time-series $T1$ and $T2$ of length $n$:

\begin{equation}
ED(T1,T2) = \sum_{i=1}^n \sqrt {(T1_i - T2_i )^2 }    
\end{equation}
The advantage of ED is its computational simplicity. However, the sensor signals in our dataset incur distortions such as acceleration and deceleration, which lock-step methods fail to account for. \textbf{Elastic measures} account for these distortions. \textbf{\textit{Dynamic Time Warping (DTW)}} is one such measure \cite{berndt1994using}; it develops a non-linear mapping to align the series. ED is a special case of DTW that allows one-to-one mapping only, while DTW finds the best alignment distance by aligning the two mappings in several ways. The $DTW$ distance between two time-series $T1$ and $T2$ of length $n$ and $m$ is:
\begin{figure}

\centering
    \includegraphics[width=0.8\linewidth]{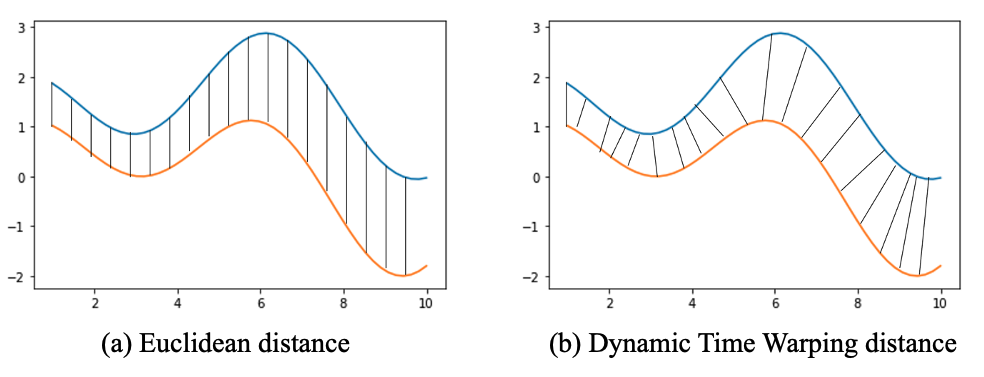}
    \caption{Comparison of dissimilarity measures.}\label{Fig_tsc_DismMsres}
\end{figure}

       



\begin{equation}
DTW(T1,T2) = \min_\pi \sum_{i,j \in \pi}^n \sqrt {(T1_i - T2_j )^2 }    
\end{equation}

where $\pi = [\pi_1,\pi_2 \dots \pi_k]$ is the warping path that defines possible series alignment. 
See figure \ref{Fig_tsc_DismMsres} for the contrast between the two measurements. A significant shortcoming of DTW is the computational cost incurred while searching for a good alignment. Warping path restriction is a typical optimization method. We used DTW with \textbf{\textit{Sakoe-Chiba }}optimization \cite{sakoe1978dynamic} to reduce the amount of off-diagonal points for alignment. 

\begin{figure}
\centering
    \includegraphics[width=\linewidth]{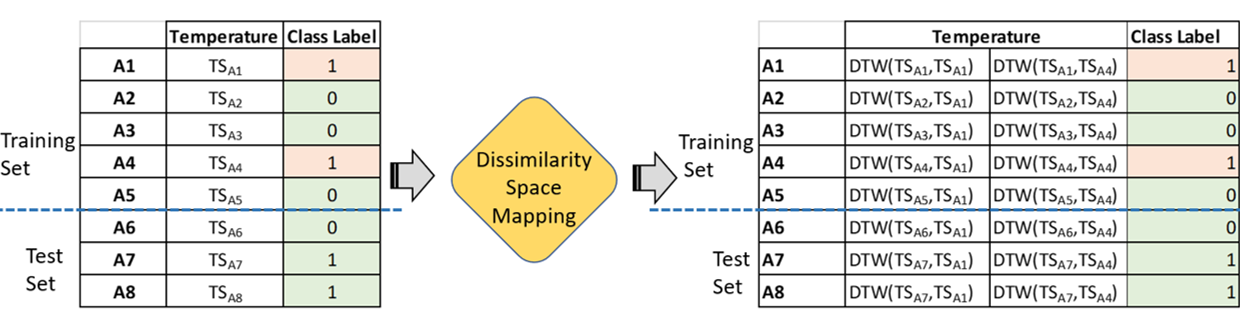}
    \caption{Dissimilarity space mapping for a univariate dataset.}\label{c3_Fig_tsc_DissimilaritySpaceMapping}
\end{figure}
\subsection{Distance features}

Distance-feature-based TSCs calculate the proximity between each pair of series to create a dissimilarity matrix. Thus, it obtains a new representation of data using distance-based features. 
A crucial design option is whether to compute distance features using the \textbf{\textit{Global}} or \textbf{\textit{Local}} patterns like \textit{shaplets} \cite{abanda2019review}. Since pattern extraction is computationally expensive, we only computed \textbf{\textit{Global distance features}}.

\subsection{Dissimilarity space \label{c3_TSC_DSMSPACE}}

As section \ref{c3_Intro} explains, dissimilarity measurements can retain complicated patterns, making them unsuitable for vector-based learning strategies. To overcome this hurdle, data is initially transformed into a dissimilarity space, after which any conventional machine-learning method can be deployed. This approach combines structural representation and statistical learning. The fundamental step in computing dissimilarity space is selecting $k$ prototype time-series. A time-series' dissimilarity representation for $k$ prototypes will have $k$ features, each representing its distance from a prototype. Thus, dissimilarity representations are vectors in $\mathbb{R}^k$ space. Let $\Upsilon$ be a time-series space, having a subset $\mathbb{P}$ of $k$ prototypes, such that


\begin{equation*}
    \mathbb{P} = \{\rho_1,\rho_2 \dots \rho_k\} \subseteq \Upsilon
\end{equation*}
This dissimilarity space of time-series $T$ with respect to $\mathbb{P}$ can be computed as:
\begin{equation*}
    \phi : \Upsilon \longrightarrow \mathbb{R}^k , T \mapsto (d(T,\rho_1),d(T,\rho_2) \dots d(T,\rho_k))
\end{equation*}
Where $d$ measures dissimilarity. In this case, the $i^{th}$ feature of $\phi(T)$ represents the distance between $T$ and the prototype $\rho_i$. Any traditional machine learning algorithm can use $\phi(T)$, the dissimilarity representation of $T$. 
Figure \ref{c3_Fig_tsc_DissimilaritySpaceMapping} illustrates dissimilarity mapping for a univariate dataset, explained as follows. The only variable here is temperature, and $\Upsilon=\{TS_{A1}, TS_{A2} \dots TS_{A8}\}$ is the entire dataset consisting of 8 time-series signals. $\mathbb{P}$ constitutes all the training set instances having class label = 1, i.e. $\mathbb{P} = \{A1, A4\} \subseteq \Upsilon$, where $|\mathbb{P}| = 2$. The resulting dissimilarity space representation has vectors in $\mathbb{R}^2$ and can be computed by calculating the distance of each time-series signal in $\Upsilon$ from $TS_{A1}$ and $TS_{A4}$.

\section{Proposed System \label{c3_Proposed_System}}

In our proposed system, we detected event-spoofing threats using distance-based TSC. We are assuming the same threat model as explained in \cite{birnbach2019peeves}. We aim to learn the correlation between IoT events and sensor time-series signals. A crucial aspect is the length and duration of each time-series signal, known as the Event Signature Window (ESW). 
To learn ESW, we utilized Relative Mutual Information (RMI) in the same way as used by Birnbach et al. \cite{birnbach2019peeves}. For each sensor, sensor window $S_{[t^-,t^+]}$ with maximum RMI is selected. Moreover, only sensors having RMI above a given \textit{\textbf{Threshold}} are selected as features to build a model for a given IoT event. Figure \ref{c3_Fig_tsc_Architecture} depicts the overall system architecture and is explained below:

\begin{figure}

\centering
    \includegraphics[width=1\linewidth]{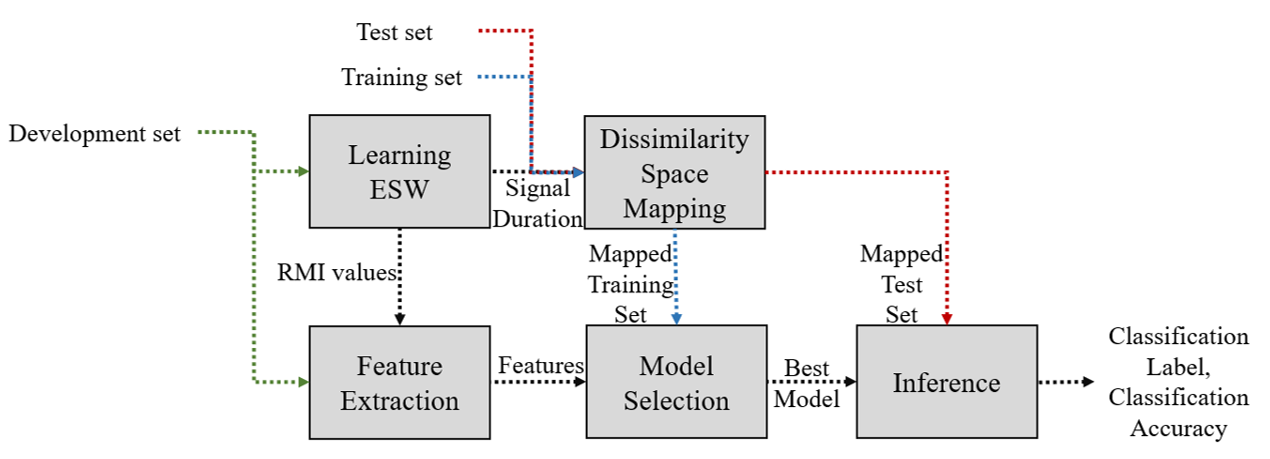}
    \caption{System Architecture.}\label{c3_Fig_tsc_Architecture}
\end{figure}
\subsection{Constructing Dissimilarity space\label{c3_distant_matrix_computation}}

Significant processing overhead is associated with dissimilarity space computation for large sets of $k$ prototypes, as explained in \ref{c3_TSC_DSMSPACE}. On the other hand, a small set may result in inadequate pattern representation. Selecting all training sets is one method. The PEEVES dataset has almost half a million training cases. Additionally, the dataset is multivariate. With such a large prototype set, dissimilarity space computation will take exponential time and require massive storage. The dataset's substantial imbalance helped us solve this problem. Since the dataset has few 1-events, we chose them as the prototype. The dissimilarity space computation overhead was greatly reduced while maintaining structural patterns.

Let $\mathbb{D}$ be the time-series dataset, divided into training set $\mathbb{D}_{train}$ and test set $\mathbb{D}_{test}$. The set of prototypes $\mathbb{P}$ is given as:
{\small
\begin{equation*}
    \mathbb{P} = \{\rho \mid event(\rho) = 1 \wedge \rho \subseteq \mathbb{D}_{train}\}
\end{equation*}
}
We employed DTW for dissimilarity measures, as explained in section \ref{c3_TSC_DisMsr}. Thus, the dissimilarity space of any time-series $T \in \mathbb{D}$ with respect to $\mathbb{P}$ is calculated as:
{\small
\begin{equation*}
    \phi : \Upsilon \longrightarrow \mathbb{R}^k ,T \mapsto (DTW(T,\rho_1),DTW(T,\rho_2) \dots DTW(T,\rho_k))
\end{equation*}
}


\subsection{Model selection\label{c3_ML}}
 After mapping features to dissimilarity space, machine learning models are easy to design. However, due to the highly unbalanced dataset with few 1-event examples, a machine learning classifier may overfit due to a low confidence level. For example, the Camera-On event has 170 positive instances and 1,050,892 negative cases (measured every second for 13 days). This imbalance could cause overfitting. Several strategies can be employed to address the overfitting issue. Adding more data can be one strategy. Previous research \cite{birnbach2019peeves} samples the dataset every second. We downsampled the dataset every 100th second; adding more data can fix the overfitting. However, adding data would only add 0-events, further unbalancing the dataset. Secondly, additional data increases computing costs. For instance, RMI-based feature selection computes multiple statistics for each possible ESW. It is computationally costly as it takes $O(n^2)$ trips through the dataset, where $n$ is the number of distinct time intervals between $t^{-}$ and $t^{+}$. For $[-30,+30]$ seconds event window search space, 1830 dataset passes are needed. A more practical approach is to explore simpler learning models, such as linear SVM, which have a strong bias and resist overfitting. Therefore, we investigated variants of three state-of-the-art classifier models, namely, Random Forest (RF), Support Vector Machine (SVM), and K-Nearest Neighbor (KNN), and chose the one with the lowest training set error to solve overfitting.

\begin{figure}[ht!]

\centering
    \includegraphics[width=0.85\linewidth]{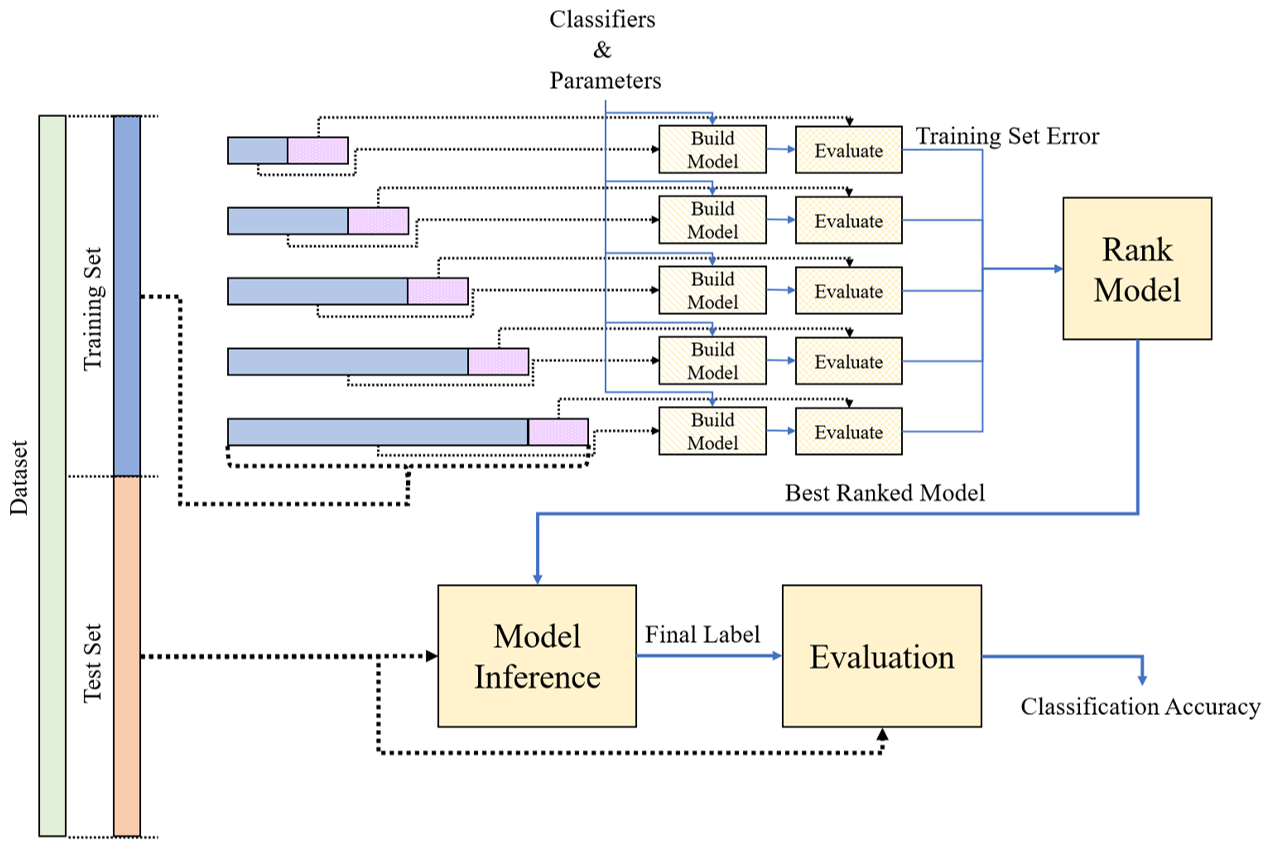}
    \caption{Model Selection.}\label{c3_Fig_tsc_ModelRank}
\end{figure}

To provide a realistic evaluation, we only use the training set error to select the model and evaluate it on a separate test set. We now discuss the ranking criteria derived from the training set errors. For each IoT event $I$, we ranked all the different classifier invariants and selected the best model (see figure \ref{c3_Fig_tsc_ModelRank}). The ranking is based on validation set error computed as follows: 
\begin{enumerate}
    \item 

    Each classifier $C \in \mathbb{C}$: set of all classifiers, cross-validation is performed. 
    Time Series Split Cross-Validation, a rolling-over KFold, is used, where in the $K^{th}$ split, the first $K$ folds train the model, while $K+1$ serves as validation. Each split's training set is a superset of the preceding splits'.

    \item The validation set from each split is used for evaluation and the rank is calculated on the bases of Average (Avg) and Standard Deviation of Equal Error Rate (EER),  Detection Rate (DR), and False Alarm Rate (FAR).
    \item Classifier rank $rank(C)$ is calculated which prefers lower EER and FAR but higher DR.
    \item For any given IoT event $I$, the best classifier $\mathbb{C}_{Best}^I$ is selected as the one having highest rank: $\max_{c \in \mathbb{C}}(rank(C))$
\end{enumerate}

\section{Evaluation \label{c3_Evaluation}}
\subsection{Experimental Setup}
First-day data from PEEVES was used as development, while the rest were evenly divided as train and test data. 
PEEVES data contains 22 IoT events. We found no 1-event in the development set data when computing ESW for Windows open and close events. Thus, we excluded these two events from the evaluation.

In their original investigation, Birnbach et al. \cite{birnbach2019peeves} sampled 0-events every second, while 1-events are rare. The "Camera on" and "Camera off" events have 70 and 69 1-events, respectively, but nearly a million 0-events, resulting in a very imbalanced data set that causes overfitting. We sampled 0-events less frequently to remedy this. This greatly decreased dataset sizes. Events were sampled every 500 seconds in the initial phase (see Table \ref{tab_TSC_500}). However, for events like Radiator, such low sampling frequency resulted in much lower accuracy. Thus, we sampled every 100 seconds to add data. The dataset was still substantially smaller than the original study. Compare dataset sizes in table \ref{tab_TSC_dataset_sizes}. 


\subsection{Performance metrics \label{lbl_perfMat}}
\subsubsection{Training set size}
Due to the ongoing flux of heterogeneous devices, IoT has few high-quality datasets. Thus, classifier efficacy as a function of training set size is critical.
\subsubsection{Classification Accuracy}

The detection rate (DR) and false alarm rate (FAR) are used to assess the reliability of the event spoofing attack classification system. Spoofing detection (DR) is the percentage of correctly recognized 0-events, whereas false alarm rate (FAR) is the percentage of 1-events misclassified as 0. Detecting fake events and minimizing the false alarm rate (FAR) is equally crucial for the classification process. Therefore, instead of a binary classifier, a probabilistic classifier is used to aggregate these two matrices. The probabilistic classifier's decision limit is dynamically set at the Equal Error Rate (EER) threshold, where DR is maximized and FAR is minimized: $EER: DR ~= 1 - FAR$.

\begin{table}[!htbp]\centering
\caption{Comparison of dataset sizes sampled at various frequencies}\label{tab_TSC_dataset_sizes}
\scriptsize
\begin{tabular}{|l|c|}\toprule
\textbf{Sampling frquency} &\textbf{Resulting dataset size}  \\\midrule
Sample every second (PEEVES)  & 1123200 \\
Sample every 100-second & 11232\\
Sample every 500-second & 2247\\

\bottomrule
\end{tabular}
\end{table}

\subsection{Results}
\begin{table}[!htbp]\centering
\caption{Time-series classification using training data {\textbf{500-times}} smaller than PEEVES (Best Classifier: RF)}\label{tab_TSC_500}
\scriptsize
\begin{tabular}{l|lll|llll}\toprule
&\multicolumn{3}{c}{\textbf{PEEVES}} &\multicolumn{4}{c}{\textbf{Our Results}} \\\cmidrule{2-8}
\textbf{Event} &\textbf{EER} &\textbf{DR} &\textbf{FAR } &\textbf{EER} &\textbf{DR} &\textbf{FAR} &\textbf{} \\\cmidrule{1-8}
Fan off &0.17\% &99.82\% &0.0\% &1.40\% &98.1\% &4.2\% & \\
Screen on &0.0\% &100\% &0.0\% &0.0\% &100\% &2.7\% & \\
Radiator on &0.97\% &98.93\% &5.26\% &\cellcolor[HTML]{fff2cc}32.5\% &\cellcolor[HTML]{fff2cc}59.6\% &\cellcolor[HTML]{fff2cc}100\% & \\
\bottomrule
\end{tabular}
\end{table}
\begin{table}[!htbp]\centering
\caption{Time-series classification using training data {\textbf{100-times}} smaller than PEEVES.(Best Classifier: SVM for \{PC on, Radiator off, Radiator on\} and RF for rest )}\label{tab_TSC_100}
\scriptsize
\begin{tabular}{l|ccc|ccc}\toprule
&\multicolumn{3}{c}{\textbf{PEEVES}} &\multicolumn{3}{c}{\textbf{Our Results}} \\\cmidrule{2-7}
\textbf{Event} &\textbf{EER} &\textbf{DR} &\textbf{FAR } &\textbf{EER} &\textbf{DR} &\textbf{FAR } \\\cmidrule{1-7}
Coffee used &0.00\% &100.00\% &0.00\% &0.00\% &100.00\% &0.00\% \\
Fridge open &0.00\% &100.00\% &0.00\% &0.00\% &100.00\% &0.00\% \\
Fridge close &0.00\% &100.00\% &0.00\% &0.00\% &100.00\% &0.00\% \\
Light off &0.00\% &100.00\% &0.00\% &0.00\% &100.00\% &0.00\% \\
Light on &0.02\% &99.98\% &0.00\% &\cellcolor[HTML]{a9d08e}0.00\% &\cellcolor[HTML]{a9d08e}100.00\% &\cellcolor[HTML]{a9d08e}0.00\% \\
PC on &0.00\% &100.00\% &0.00\% &0.00\% &100.00\% &0.00\% \\
Screen on &0.00\% &100.00\% &0.00\% &0.00\% &100.00\% &0.00\% \\
Screen off &0.29\% &99.71\% &0.00\% &9.09\% &100.00\% &9.09\% \\
Camera on &1.00\% &99.99\% &0.00\% &0.00\% &99.92\% &0.00\% \\
Radiator off &49.41\% &7.65\% &100.00\% &\cellcolor[HTML]{a9d08e}0.00\% &\cellcolor[HTML]{a9d08e}94.20\% &\cellcolor[HTML]{a9d08e}87.50\% \\
Radiator on &\cellcolor[HTML]{b4c6e7}0.97\% &\cellcolor[HTML]{b4c6e7}98.93\% &\cellcolor[HTML]{b4c6e7}5.26\% &22.22\% &0.00\% &66.67\% \\
Doorbell &15.06\% &0.36\% &55.56\% &\cellcolor[HTML]{a9d08e}0.00\% &\cellcolor[HTML]{a9d08e}96.94\% &\cellcolor[HTML]{a9d08e}21.43\% \\
Shade down &\cellcolor[HTML]{b4c6e7}11.39\% &\cellcolor[HTML]{b4c6e7}0.28\% &\cellcolor[HTML]{b4c6e7}15.15\% &43.75\% &0.00\% &81.25\% \\
Shade up &\cellcolor[HTML]{b4c6e7}12.04\% &\cellcolor[HTML]{b4c6e7}0.19\% &\cellcolor[HTML]{b4c6e7}12.50\% &31.25\% &0.00\% &68.75\% \\
Door close &0.00\% &100.00\% &0.00\% &0.00\% &99.88\% &0.00\% \\
PC off &\cellcolor[HTML]{b4c6e7}0.39\% &\cellcolor[HTML]{b4c6e7}99.61\% &\cellcolor[HTML]{b4c6e7}0.00\% &22.22\% &0.00\% &22.22\% \\
Door open &\cellcolor[HTML]{b4c6e7}0.00\% &\cellcolor[HTML]{b4c6e7}100.00\% &\cellcolor[HTML]{b4c6e7}0.00\% &12.35\% &0.00\% &20.99\% \\
Fan on &\cellcolor[HTML]{b4c6e7}0.00\% &\cellcolor[HTML]{b4c6e7}100.00\% &\cellcolor[HTML]{b4c6e7}0.00\% &22.97\% &0.19\% &23.00\% \\
Camera off &\cellcolor[HTML]{b4c6e7}1.00\% &\cellcolor[HTML]{b4c6e7}99.90\% &\cellcolor[HTML]{b4c6e7}0.00\% &27.80\% &44.80\% &97.20\% \\

\bottomrule
\end{tabular}
\end{table}


We first attempted with few events with a further reduced dataset, where events are sampled every 500-sec (see table \ref{tab_TSC_500}). For Fan off and Screen on, we achieved the same accuracy as that by PEEVES, despite using an incredibly smaller dataset. 
For Radiator on, the accuracy was relatively poor. In the case of the Radiator on event, our classifier experienced over-fitting, therefore, we decided to add back data and sampled data after a 100-sec interval. The results are presented in table \ref{tab_TSC_500}.

As shown, we can accurately predict most events like PEEVES. Our approach sometimes outperformed PEEVES, but sometimes PEEVES did better. See highlighted cases in table \ref{tab_TSC_100} and summarised below:
\begin{itemize}
    \item Screen on and Radiator on events performed better with more data. 
    \item Our method improved two of PEEVES' six worst events, Radiator on and Doorbell used, using a much smaller dataset. 
    \item Camera off, PC on, Fan on, and Door open events severely hampered our classifier. However, their counter events, i.e., Camera on, PC off, Fan off, and Door close demonstrated perfect accuracy. This led us to believe that the RMI values learned were likely inaccurate for these events.
\end{itemize}

\begin{table}[!htbp]\centering
\caption{Door open off event accuracy using \textbf{\textit{End-to-End TSC}}  (Best Classifier: Random Forest) }\label{tab_TSC_DoorOpen}
\scriptsize
\begin{tabular}{l|lll|llll}\toprule
&\multicolumn{3}{l}{\textbf{PEEVES}} &\multicolumn{4}{l}{\textbf{Our Results}} \\\cmidrule{2-8}
\textbf{Event} &\textbf{EER} &\textbf{DR} &\textbf{FAR } &\textbf{EER} &\textbf{DR} &\textbf{FAR } & \\\cmidrule{1-8}
Door open &0.00\% &100.00\% &0.00\% &0.00\% &100.00\% &0.00\% &\\
\bottomrule
\end{tabular}
\end{table}

Finally, we decided to deploy an \textbf{\textit{End-to-End TSC}} solution. This meant utilizing distance-based transformation for all three phases of the system, namely feature extraction, learning ESW, and classification. This model is evaluated on one of the poorest performing event class: Door open. As seen in table \ref{tab_TSC_DoorOpen}, we were able to achieve 100\% accuracy, although we used the reduced dataset size sampled every 500-sec. We intend to test this model for all events in our future work.

\section{Conclusion \& Future work\label{c3_summary}}
This paper presents a smart home event-spoofing detection method using time-series classification (TSC). The data is time-series with distinct temporal patterns. Traditional statistical feature extraction methods lose these features, resulting in poor learning. We present a novel Time Series Classification method to detect these attacks. TSC is used by mapping the time-series dataset to a dissimilarity space using Dynamic Time Warping distance measurements. We tested 93 classifier versions to reduce model overfitting in the unbalanced dataset. These classifiers are ranked by the least training set error. The proposed system is thoroughly evaluated and benchmarked against publicly available datasets. The effectiveness of the proposed system is evident from the fact that it can achieve comparable results despite using a tremendously smaller dataset, making it suitable for the IoT ecosystem.





\bibliographystyle{IEEEtran}
\bibliography{references}

\end{document}